\documentclass[a4,12pt]{elsarticle}
\usepackage{graphicx}
\usepackage{amsmath}
\usepackage{amssymb}
\usepackage{amsthm}
\usepackage{amsfonts}
\begin{document}
\begin{frontmatter}
\title{Nonlinear optical properties of a three-dimensional anisotropic quantum dot }
\author{Tairong Chen$^{\ast}$\footnote{E-mail: Chen-tairong4487@hotmail.com\\Tel: +8615013005504 } \quad Wenfang Xie}
\address{(Department of Physics, College of Physics and
Electronic Engineering, Guangzhou University, Guangzhou 510006, PR China)}
\begin{abstract}
The linear and nonlinear optical properties in a three-dimensional anisotropic quantum dot subjected to a uniform magnetic field directed with respect to the $z-$axis have been investigated within the compact-density matrix formalism and the iterative method. The dependence of the linear and nonlinear optical properties on characteristic frequency of the parabolic potential, on the magnetic field and on the incident optical intensity is detailedly studied. Moreover, take into account the position-dependent effective mass, the dependence of the linear and nonlinear optical properties on the dot radius is investigated. The results show that the optical absorption coefficients (ACs) and refractive index (RI) changes of the anisotropic quantum dot (QD) have strongly affected by these factors, and the position effect also plays an important role in the optical ACs and RI changes of the anisotropic QD.
\end{abstract}
\begin{keyword}
A. Anisotropic QD\sep D. Absorption coefficient \sep D. Refractive index
\end{keyword}
\end{frontmatter}
\section{\textbf{Introduction}}
With the immense advance of modern nanotechnology these decades, it has become possible to confine the charge carriers in nanostructures, such as quantum dots (QDs), quantum wires, quantum wells and superlattices, with well controlled shape size and other properties [1-5]. Since QDs are created mainly by imposing a lateral confinement potential to electrons in a very narrow quantum well, the features which should be mainly considered in a QD are geometric shape, size and confining potential. As we know the physical properties can not be extrapolated from behavior at large sizes because of the prevalence of quantum effects. So the investigation of QDs has attracted enormous theoretical and experimental interest [6-10]. Further, the distinctive electrons and optical properties of QDs show a considerable potential application in electronic optoelectronic device.

It is especially necessary to investigate the optical properties of low-dimensional semiconductor because the study of optical transitions yields important information about the energy spectrum, the Fermi surface of electrons, and the value of the electronic effective mass. A great deal of work has been done on the linear and nonlinear optical properties of low-dimensional semiconductor systems, and numerous of important new information is discovered. For instance, nonlinear optical properties of an off-center donor in a QD in presence of applied magnetic field is investigated by Wenfang Xie [11], the results show the ion position, the applied magnetic field, the confinement strength and the incident have an important influence on the nonlinear optical properties of off-center donors. Therefore, some special needs can be met by modulating these physical quantities in designing electro-optical devices. In addition, the researches on isotropic parabolic potential have drawn great attention, and lots of corresponding reports have been published. Also the effects of the parabolic on the energy and optical properties have been widely investigated. In many studies, an isotropic harmonic oscillator potential are used as the lateral confinement of electrons to do the optical research for the reasons that the parabolic confinement is more appropriate when the QD is fabricated and the parabolic gives a quadratic Hamiltonian, the spectrum of which can be obtained by a purely algebraic method. As far as we know, a number of important optical information has been obtained and many nonlinear problems have been resolved by this method [12-14].

In recent years, the theory for boundary value problems for anisotropic quantum systems have aroused significant interest of the researchers because the new and useful effects have been discovered in such quantum systems [15-21]. For example, the
resonance Raman scattering in the anisotropic QDs in the presence of arbitrarily directed magnetic field
and polarization vector have theoretically investigated by Shorokhov and Margulis [15], the results reveal that the QD subjected in a magnetic field can be used as the detector of phonon
modes. Hybrid resonances in the optical absorption of a three-dimensional anisotropic quantum well has been studied by Geyler $et$ $ al.$ [16], the results show that in the general case there are three resonance frequencies on the absorption curve, and in the limit $T\rightarrow0$ ( the case of a degenerate gas), the absorption
curve as a function of the radiation frequency contains kinks. In addition, hybrid-impurity resonances in anisotropic QDs has been discussed by Margulis $et$ $ al.$ [17], and they reported that the scattering of electrons by impurities leads to the resonance absorption even if we have
only one impurity in the QD; the position of the hybrid-impurity resonances depends strongly on the magnetic field and the
characteristic frequencies of the parabolic confinement. Moreover, magnetoexcitons in quantum wires with an anisotropic parabolic potential has been calculated by Tanaka $et$ $ al.$ [18]. Their results show that the experimental results on magnetophotoluminescence (magneto-PL) of GaAs QWR's can be well explained by this model, demonstrating the importance of the exciton effect to understand the magneto-PL properties of QWR's. All these results indicate that the optical properties are strongly dependent on the anisotropy
of the material structure.

In this paper, we present a theoretical study of the linear and nonlinear optical absorption coefficients (ACs) and refractive index (RI) changes in a three-dimensional QD subjected to a uniform magnetic field directed with respect to the $z-$axis. To simplify the calculations, the linear canonical transformations of the phase space has been adopted [16,17]. By means of these transformations, one can find a new coordinate system such that in the new phase coordinates the wave functions have the simplest form.
From the consideration of device applications, engineering the electric structure of materials by means of characteristic frequency, magnetic field, incident optical intensity and the dot radius, offers the possibility of tailoring the energy spectrum to produce desirable nonlinear optical properties.
\section{\textbf{Model and theory}}
\subsection{Electronic state in an anisotropic quantum dot }
Within the framework of the effective mass approximation,
the Hamiltonian description of an electron confined in an anisotropic parabolic QD in the presence of magnetic field $\vec{B}$ has the form
\begin{equation}
H=\frac{1}{2m^*}(\vec{p}-\frac{e}{c}\vec{A})^2+\frac{m^*}{2}(\Omega_{x}^2x^2+\Omega_{y}^2y^2+\Omega_{z}^2z^2),
\end{equation}
where $m^*$ is the electronic effective mass, $\vec{A}$ is the vector potential of magnetic field $\vec{B}$,
$\Omega_{i}$ $(i=x,y,z)$ are the characteristic frequencies of parabolic potential. Here we choose the following gauge for the vector potential:
\begin{equation}
\vec{A}=(\frac{1}{2}B_{y}z-B_{z}y,0,B_{x}y-\frac{1}{2}B_{y}x).
\end{equation}
It is surely a complicate problem to calculate directly the matrix elements of electron-photon interaction. To resolve this problem, a method of canonic transformation of the phase
space has been adopted via using only simple
calculations from linear algebra, and which has been successfully used to resolve many problems about the anisotropic systems [16,18-21]. Within the linear canonic
transformation of the phases pace, the Hamiltonian can be represented with canonic form in the new phase
coordinates ($\vec{P}, \vec{Q}$)
\begin{equation}
H(\vec{P},\vec{Q})=\frac{1}{2m^*}(P_{1}^2+P_{2}^2+P_{3}^2)+\frac{m^*}{2}(\omega_{1}^2Q_{1}^2+
\omega_{2}^2Q_{2}^2+\omega_{3}^2Q_{3}^2),
\end{equation}
where $\omega_{i}$ $(i=1, 2, 3)$ are the hybrid frequencies related to the magnitude and direction of the magnetic field $\vec{B}$, and the frequencies are obtained from six-order algebraic equation [22]. Hence,
the spectrum of Hamiltonian can be obtained as
\begin{equation}
E_{nml}=\hbar\omega_{1}(n+\frac{1}{2})+\hbar\omega_{2}(m+\frac{1}{2})+\hbar\omega_{3}(l+\frac{1}{2}),
 \quad n,m,l,=0, 1, 2,\ldots
\end{equation}
and the corresponding wave functions in the new phase coordinates have the form
\begin{equation}
\Phi_{nml}(Q_{1},Q_{2},Q_{3})=\phi_{n}(Q_{1})\phi_{m}(Q_{2})\phi_{l}(Q_{3})
\end{equation}
where $\phi_{j}(Q)$ are the oscillator functions, and $j=n,m,l$.\\
\subsection{Calculation of linear and third-order nonlinear optical absorption coefficients and refractive index changes}
In the present work, the compact-density approach to calculate absorption coefficient and refractive index for a three-dimensional anisotropic QD. For simplicity, suppose the system is excited by a monochromatic electromagnetic field as
\begin{equation}
E(t)=E_{0}\cos{(\omega t)}=\widetilde{E}e^{i\omega t} +\widetilde{E}e^{-i\omega t},
\end{equation}
where $\omega$ is the frequency of the external field incident with a polarization vector normal to the QD along $z$-direction. The time-dependent equation of the density matrix operator $\varrho$ is given by
\begin{equation}
\frac{\partial\varrho_{ij}}{\partial t}=\frac{1}{i\hbar}[H_{0}-qzE(t),\varrho]_{ij}-\Gamma_{ij}(\varrho-\varrho^{(0)})_{ij},
\end{equation}
where $H_{0}$ is the Hamiltonian of the system without the incident field $E_{(t)}$, $q$ is the electronic charge. $\varrho^{(0)}$ is the unperturbed density matrix and $\Gamma_{ij}=\frac{1}{\tau_{ij}}$ is the relaxation rate representing the damping due to the electron-phonon interaction. Using the iterative method, the time-dependent density matrix operator can be solved as
\begin{equation}
\varrho(t)=\sum_{n}\varrho^{(n)}(t),
\end{equation}
with
\begin{equation}
\frac{\partial\varrho_{ij}^{(n+1)}}
{\partial t}=\frac{1}{i\hbar}\{[H_{0},\varrho^{(n+1)}]_{ij}-i\hbar\Gamma_{ij}\varrho^{(n+1)}_{ij}\}
\frac{1}{i\hbar}[qz-\varrho^{(n)}]_{ij}E(t).
\end{equation}
The electronic polarization $\vec{P}(t)$ and susceptibility $\chi(t)$ are defined by the dipole operator $M$, and the density matrix $\varrho$, respectively,
\begin{equation}
\vec{P}(t)=\epsilon_{0}\chi_{\omega}\widetilde{E}e^{i\omega t}+\epsilon_{0}\chi_{-\omega}\widetilde{E}e^{-i\omega t}
=\frac{1}{V}Tr(\varrho M),
\end{equation}
where $\epsilon_{0}$ is the permittivity of free space, $V$ is volume of the system. $Tr$ denotes the trace or the summation over the diagonal elements of the matrix $\varrho M$, the analytic expressions of the linear and third-order nonlinear susceptibilities can be obtained as follows:
for the linear term
\begin{equation}
\epsilon_{0}\chi^{(1)}(\omega) =\frac{\rho|M_{fi}|^2}{E_{fi}-\hbar\omega-i\hbar\Gamma_{fi}},
\end{equation}
and for the third-order nonlinear term
\begin{eqnarray}
\epsilon_{0}\chi^{(3)}(\omega)&=&-\frac{\rho|M_{fi}|^2\widetilde{E}^2}{E_{fi}-\hbar\omega-i\hbar\Gamma_{if}}
[\frac{4|M_{fi}|^2}{(E_{fi}-\hbar\omega)^2+(\hbar\Gamma_{fi})^2}\nonumber\\
&-&\frac{(M_{ff}-M_{ii})^2}{(E_{fi}-i\hbar\Gamma_{fi})(E_{fi}-\hbar\omega-i\hbar\Gamma_{if})}],
\end{eqnarray}
where $\rho$ is carriers density. $E_{fi}=E_{f}-E_{i}$ is the energy interval of the two level system. $M_{fi}=e\langle \Phi_{f}|x|\Phi_{i}\rangle$ denotes the electric dipole moment of the transition from $\Phi_{i}$ state to $\Phi_{f}$ state. $\Gamma$ is phenomenological operator. Non-diagonal matrix element $\Gamma_{if}$ $(i\neq f)$ of operator $\Gamma$, which is called as relaxation rate of $f$th state, is the inverse of the relaxation time $\tau_{if}$. The susceptibility $\chi(\omega)$ is related to the absorption coefficient $\alpha(\omega)$ and the changes in the refractive index $\Delta n(\omega)/n_{r}$ as follows
\begin{equation}
\alpha(\omega)=\omega\sqrt{\frac{\mu}{\epsilon_{R}}}Im(\epsilon_{0}\chi(w)),
\end{equation}
\begin{equation}
\frac{\Delta n(\omega)}{n_{r}}=Re(\frac{\chi(\omega)}{2n_{r}}),
\end{equation}
where $\mu$ is the permeability of the material, $\epsilon_{R}=n_{r}^2\epsilon_{0}$ is the real part of the permittivity.

The linear and third-order nonlinear absorption coefficients are obtained as
\begin{equation}
\alpha^{(1)}(\omega)=\omega\sqrt{\frac{\mu}{\epsilon_{R}}}\frac{\rho|M_{fi}|^2\hbar\Gamma_{if}}
{(E_{fi}-\hbar\omega)^2+(\hbar\Gamma_{fi})^2},
\end{equation}
\begin{eqnarray}
\alpha^{(3)}(\omega,I)&=&-\omega\sqrt{\frac{\mu}{\epsilon_{R}}}(\frac{I}{2\epsilon_{0}n_{r}c})
\frac{\rho|M_{fi}|^2\hbar\Gamma_{if}}
{(E_{fi}-\hbar\omega)^2+(\hbar\Gamma_{fi})^2}[4|M_{fi}|^2\nonumber\\
&-&\frac{|M_{ff}-M_{ii}|^2[3E_{fi}^2-4E_{fi}\hbar\omega+\hbar^2(\omega^2-\Gamma_{if}^2)]}{E_{fi}^2+
(\hbar\omega_{if})^2}].
\end{eqnarray}
Here $I$ is the intensity of incident radiation, and the total absorption coefficient $\alpha(\omega,I)$ is given by
\begin{equation}
\alpha(\omega,I)=\alpha^{(1)}(\omega)+\alpha^{(3)}(\omega,I).
\end{equation}
The linear and third-order nonlinear refractive index changes are given by
\begin{equation}
\frac{\Delta n^{(1)}(\omega)}{n_{r}}=\frac{\rho|M_{fi}|^2}{2n_{r}^2\epsilon_{0}}\frac{E_{fi}-\hbar\omega}
{(E_{fi}-\hbar\omega)^2+(\hbar\Gamma_{fi})^2},
\end{equation}
and
\begin{eqnarray}
\frac{\Delta n^{(3)}(\omega)}{n_{r}}&=&-\frac{\rho|M_{fi}|^2}{4n_{r}^3\epsilon_{0}}
\frac{\mu c I}{[(E_{fi}-\hbar\omega)^2+(\hbar\Gamma_{fi})^2]^2}\nonumber\\
& &\times[4(E_{fi}-\hbar\omega)|M_{fi}|^2-\frac{|M_{ff}-M_{ii}|^2}{E_{fi}^2+(\hbar\omega)^2}
((E_{fi}-\hbar\omega)\nonumber\\
& &[E_{fi}(E_{fi}-\hbar\omega)-(\hbar\Gamma_{if})^2]-(\hbar\Gamma_{if})^2(2E_{fi}-\hbar\Gamma_{if}))].
\end{eqnarray}
Therefore, the total refractive index change can be obtained as
\begin{equation}
\frac{\Delta n(\omega)}{n_{r}}=\frac{\Delta n^{(1)}(\omega)}{n_{r}}+\frac{\Delta n^{(3)}(\omega)}{n_{r}}.
\end{equation}
\section{\textbf{Result and Discussions}}
In this study, the linear, the third-order nonlinear and the total ACs and RI changes of the three-dimensional anisotropic $GaAs/AlGaAs$ QD subjected to the uniform magnetic field directed with respect to the $z-$axis have been numerically investigated. The physical parameters for calculations are used as follows [23-25]: $\rho=5.0\times10^{22} m^{-3}$, $\tau=0.14 s$, $n_{r}=3.2$ and $m^*=0.067 m_{0}$, where $m_{0}$ is the free electronic mass. The other two characteristic frequencies are set as $\Omega_{y}=2.0\times10^{12}s^{-1}$ and $\Omega_{z}=1.5\times10^{11} s^{-1}$. The results of our calculations are presented in figures 1-8. Fig. 1 presents the linear, the third-order nonlinear and the total ACs of a three-dimensional anisotropic QD as a function of the incident photon energy $\hbar\omega$ for three different values of characteristic frequency $\Omega_{x}=0.9\times10^{13} s^{-1}$, $1.1\times10^{13} s^{-1}$ and $1.3\times10^{13} s^{-1}$, respectively. In Fig. 1, the applied magnetic field $B$ is set to be $1.0 T$, and the incident optical intensity is set to be $1.0 Mw/cm^{2}$. From Fig. 1, it can be easily found that the ACs as a function of $\hbar\omega$ in the three-dimensional anisotropic parabolic potential QD share the similar physical features with that in the isotropic parabolic potential QD [26]. Moreover, it can be noted that the linear, third-order nonlinear and total ACs are not monotonic functions of $\hbar\omega$. For each $\Omega_{x}$, the linear, third-order nonlinear, and total optical ACs as a function of $\hbar\omega$ has a prominent peak, respectively, at some position because of the one-photon resonance enhancement. And when the magnitude of $\hbar\omega$ is larger than $30 meV$, all the ACs will be extremely slow approaching zero as $\hbar\omega$ further increases. Additionally, it is can be found the linear ACs are monotonic functions of the characteristic frequency $\Omega_{x}$, and the values of which are monotonously increased as $\Omega_{x}$ increases. While the nonlinear terms are monotonously decreased due to the negative sign. Also, the increase of the $\Omega_{x}$ shifts the peak position of the corresponding AC to higher frequencies. It indicates a strong confinement-induced blue shift of an anisotropic parabolic resonance in semiconductor QDs. This physical behavior is originated from an augment of the energy difference between the energy levels with the enhancement of $\Omega_{x}$. So it can be concluded that the ACs of the anisotropic QD are strongly dependent on the characteristic frequency.

The RI changes are always thought to be another important parameters in optical studies of QDs. The linear, third-order nonlinear and total RI changes of the three-dimensional anisotropic QD as a function of $\hbar\omega$ are illustrated in Fig. 2 with the same parameters as taken in Fig. 1. As can be seen from this figure, the total RI changes are remarkably reduced by the third-order nonlinear RI changes because the nonlinear terms are opposite in sign of the linear ones. It is indicated that the third-order nonlinear term has an important contribution in the variation of total RI change, although the change in the total RI is mainly caused by the linear term. Moreover, one can observe that the RI changes are not monotonic functions of $\hbar\omega$. At beginning, the linear and total RI changes increase steadily with photon energy, and then reach the maximum values at some $\hbar\omega$. However, as $\hbar\omega$ approaches the threshold energy, the linear and total RI changes decrease quickly and reach the minimum values. Note that, in this region, the dispersion for any frequency of the incident photon changes its
sign. So this region is called as an absorption band due to the strong optical absorption [11]. On the other hand, it can be easily seen that, with increase of $\Omega_{x}$, the RI peaks will move to right side, which shows a blue shift of the resonance in QD. This physical behavior can be explained as that the energy interval between the initial state and the final state in an anisotropic QD becomes wider with increasing $\Omega_{x}$. Obviously, the RI changes are greatly affected by the characteristic frequency.

In Fig. 3 and 4, the linear, third-order nonlinear and total optical ACs and RI changes are displayed, respectively, as a function of the incident photon energy $\hbar\omega$ for three different values of magnetic field with $\Omega_{x}=1.2\times10^{13} s^{-1}$, and $I=1.0 Mw/cm^2$. The magnetic field effect on the optical properties has been clearly shown in these two figures. One can see that the ACs and RI changes are also not monotonic functions of $\hbar\omega$ for the different magnetic field. In addition, it is found that the ACs and RI changes as a function of $\hbar\omega$ are very sensitive to the magnetic field. Just as shown in the figures, the strong absorption can be obtained at a weak magnetic field, and the peak values of ACs are rapidly increased by the increasing $B$.
Moreover, the linear ACs and RI changes are increased with increase of $B$, while the third-order nonlinear ACs and RI changes are significantly reduced because of their negative sign. Besides, the increase of $B$ enhances the total terms and moves all total optical absorption peaks to higher energy. Note that all these behaviors are originated from the effect of the magnetic field on the anisotropic parabolic potential. The increase of applied magnetic field causes an enhancement in the confinement of electron in
the dot and leads to a larger separation between the initial energy level and the final energy level.

From Eqs. (16) and (19), it is theoretically shown that the nonlinear terms strongly depend on the incident photon intensity, so the contribution of nonlinear terms in ACs and RI changes is enhanced by an increase of incident photon intensity. In order to illustrate this effect, the linear, third-order nonlinear and total optical ACs and RI changes are plotted as a function of the incident photon energy $\hbar\omega$ for different values of incident optical intensity $I$ with $\Omega_{x}=1.2\times10^{13} s^{-1}$, and $B=1.0 T$ in the Fig. 5 and 6. As expected, the linear terms can not be changed by the variation of $I$, but the third-order nonlinear terms are monotonously decreased by the increasing $I$, where the decrease is originated from the negative sign of the third-order nonlinear terms. As a result, the total ACs and the total RI changes are reduced. Moreover, from Fig. 5, one can see that the absorption is strongly bleached at sufficiently high incident optical intensities, and the strong absorption begins to occur at around $I=2.0 Mw/cm^2$ in this work. When the incident optical intensity exceeds this value, the total absorption peak will be evidently spilt into two peaks, which is a consequence of the strong absorption. In addition, there is no shift at the resonance peak position with the variation of the incident optical intensity because the energy difference can not be affected by $I$. On the other hand, as can be seen from Fig. 6, the linear RI changes do not change with the intensity, while the third-order nonlinear terms are enhanced by increasing intensity. And that since the third-order term is opposite in sign of the linear one, any enhancement in the magnitude of nonlinear term will lead to a reduction in the difference between the linear term and the corresponding third-order nonlinear term. As a result, the total RI change is reduced. Therefore, to obtain a larger RI change, a weaker incident optical intensity should be applied. This result is in good agreement with the available refs. [10,11].

It is well know that the electron effective mass is strongly affected by its position, and the relationship between them has been reported by Peter $et$ $ al. $ [27] as
$\frac{1}{m(\vec{r})}=\frac{1}{m^*}+[1-\frac{1}{m^*}]\exp{[-\xi r]}$, where $r$ denotes the dot radius, and $\xi$
is a constant which is chosen to be $0.01 a.u.$ This choice is selected taking into account that as $r\rightarrow0$,
the particle is strongly bound within a $\delta$- function well, whereas
as $r\rightarrow\infty$, the system is three-dimensional and characterized by
the conduction band effective mass $m^*$ [27,28]. By using this expression, it can be easily explained that there
is no appreciable change of the effective mass for a large dot radius, which also shows that mass variations are unimportant for large dots. Take into account the position-dependent effective mass, the linear, the third-order nonlinear and the total optical ACs and RI changes are plotted, respectively, in Fig. 7 and 8 as a function of the incident photon energy $\hbar\omega$ for three different values of dot radius with $\Omega_{x}=1.2\times10^{13} s^{-1}$, $B=1.0 T$ and $I=1.0 Mw/cm^{2}$. From Fig. 7, it can be seen that the linear ACs are monotonously increased with the increase of dot radius $r$, as opposed to the third-order nonlinear ones. The physical origin is that the electron effective mass is strongly dependent on its position. So it is clear that the mass variation with position plays an important role in the three-dimensional anisotropic QDs.  On the other hand, from Fig. 8, it can be readily observed that the increase of dot radius leads to an increase of the maximum values and a decrease of the minimum values of the corresponding RI changes, respectively. This again shows the importance of mass
variation with position for the three-dimensional anisotropic QDs. Such a strong dependence of the linear, third-order nonlinear and total ACs and RI changes on the dot radius would be important for many device application.

\section{\textbf{Summary}}
In this paper, we have presented a detail study of the linear and nonlinear optical ACs and RI changes in a three-dimensional anisotropic QD subjected to a uniform magnetic field directed with respect to the $z-$axis within the framework of effective-mass approximation. Using density matrix theory and
iterative method, the expressions of linear and third-order nonlinear optical ACs and RI changes have been obtained. The dependence of the optical absorption and refractive index change on the characteristic frequency $\Omega_{x}$ of the parabolic potential, on the magnitude of the magnetic field $B$, on the incident optical intensity $I$, and on the dot radius $r$ has been investigated. It has been revealed that the nonlinear optical properties of the three-dimensional anisotropic QD are strongly dependent on these factors. For example, it is shown that the peak positions of linear and nonlinear terms shift to higher photon energy as the characteristic frequency increases; the ACs and RI changes will be enhanced with increasing magnetic field; the nonlinear part of ACs and RI changes are increased by increasing the incident optical intensity. And also the increase of incident optical intensity leads to a reduction of the total ACs and RI changes. Furthermore, the linear, third-order nonlinear and total ACs and RI changes are obviously affected by the dot radius.
The present results are useful for further understanding the nonlinear optical properties of a three-dimensional anisotropic QD, and we hope that this theoretical study can make a significant contribution to experimental studies and practical applications.

\section{\textbf{Acknowledgement}}
This work is supported by National Natural Science
Foundation of China (under Grant No. 11074055).

\newpage

\section{caption}

Fig. 1. The linear, the third-order nonlinear and the total ACs of a three-dimensional anisotropic QD as a function of the incident photon energy $\hbar\omega$ for three different values of characteristic frequency  $\Omega_{x}$ with $B=1.0 T$, $I=1.0 Mw/cm^2$.\\

Fig. 2. The linear, the third-order nonlinear and the total RI changes of a three-dimensional anisotropic QD as a function of the incident photon energy $\hbar\omega$ for three different values of characteristic frequency $\Omega_{x}$. Parameters are taken the same as Fig. 1.\\

Fig. 3. The linear, the third-order nonlinear and the total ACs of a three-dimensional anisotropic QD as a function of the incident photon energy $\hbar\omega$ for three different values of the applied magnetic field $B$ with $\Omega_{x}=1.2\times10^{13} s^{-1}$, $I=1.0 Mw/cm^2$.\\

Fig. 4. The linear, the third-order nonlinear and the total RI changes of a three-dimensional anisotropic QD as a function of the incident photon energy $\hbar\omega$ for three different values of the applied magnetic field $B$. Parameters are taken the same as Fig. 3.\\

Fig. 5. The linear, the third-order nonlinear and the total ACs of a three-dimensional anisotropic QD as a function of the incident photon energy $\hbar\omega$ for three different values of the incident optical intensity $I$ with $\Omega_{x}=1.2\times10^{13} s^{-1}$, $B=1.0T$.\\

Fig. 6. The linear, the third-order nonlinear and the total RI changes of a three-dimensional anisotropic QD as a function of the incident photon energy $\hbar\omega$ for three different values of the incident optical intensity $I$. Parameters are taken the same as Fig. 5.\\

Fig. 7. The linear, the third-order nonlinear and the total ACs of a three-dimensional anisotropic QD as a function of the incident photon energy $\hbar\omega$ for three different values of the dot radius $r$ with $\Omega_{x}=1.2\times10^{13} s^{-1}$, $B=1.0 T$ and $I=1.0 Mw/cm^2$.\\

\newpage
Fig. 8. The linear, the third-order nonlinear and the total RI changes of a three-dimensional anisotropic QD as a function of the incident photon energy $\hbar\omega$ for three different values of the dot radius $r$. Parameters are taken the same as Fig. 7.\\

\newpage
\begin{figure}[tbp]
\begin{center}
\includegraphics[scale=1.0]{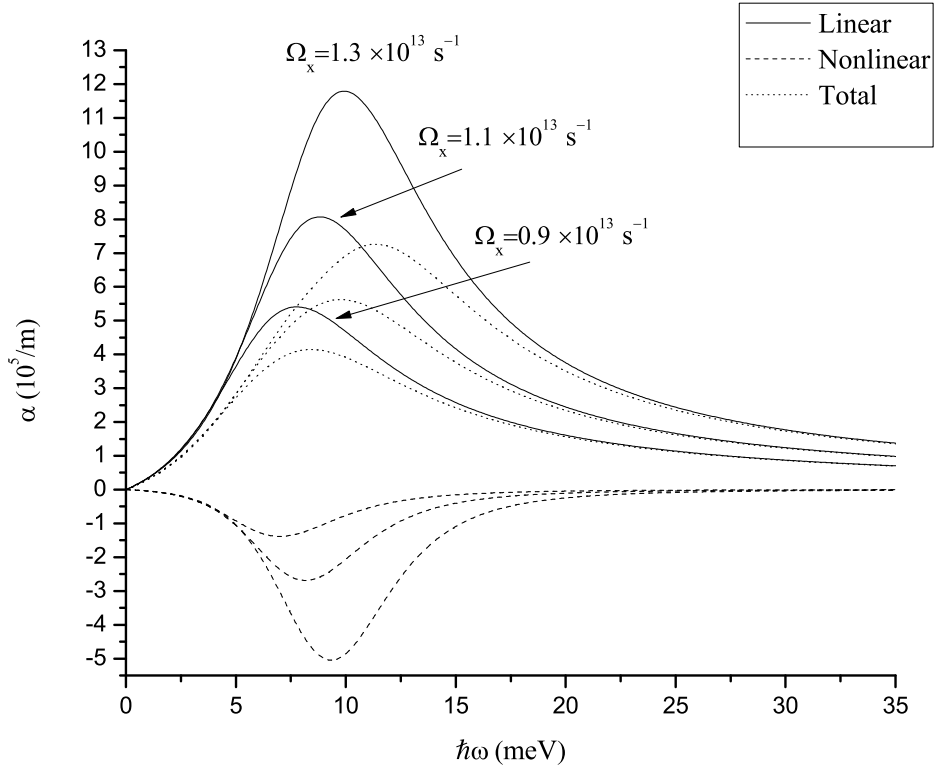}
\end{center}
\caption{The linear, the third-order nonlinear and the total ACs of a three-dimensional anisotropic QD as a function of the incident photon energy $\hbar\omega$ for three different values of characteristic frequency  $\Omega_{x}$ with $B=1.0 T$, $I=1.0 Mw/cm^2$.}
\end{figure}

\begin{figure}[tbp]
\begin{center}
\includegraphics[scale=1.0]{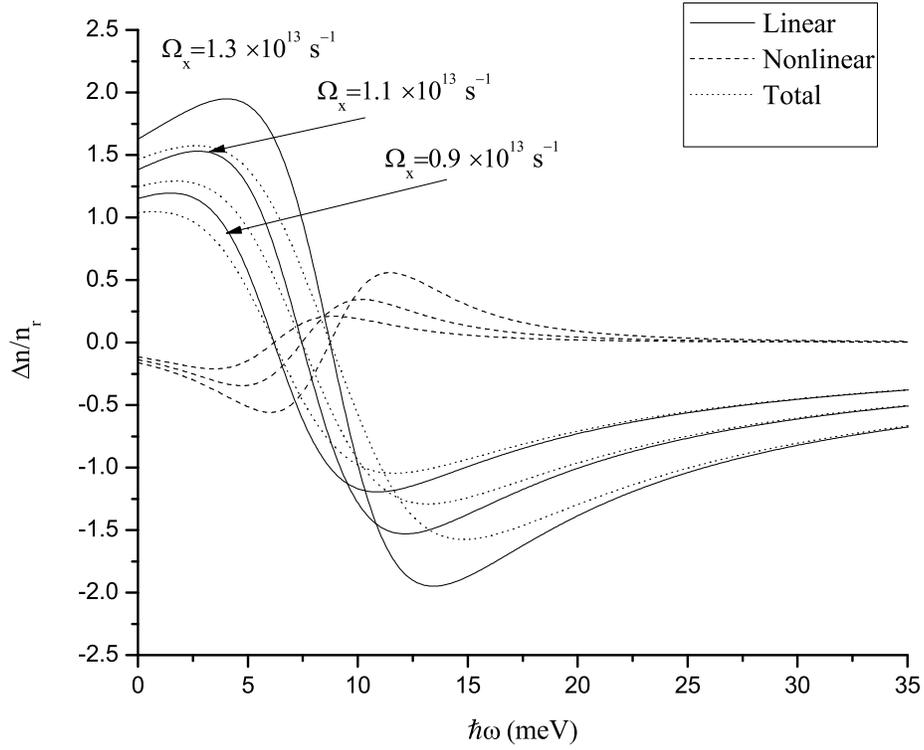}
\end{center}
\caption{The linear, the third-order nonlinear and the total RI changes of a three-dimensional anisotropic QD as a function of the incident photon energy $\hbar\omega$ for three different values of characteristic frequency $\Omega_{x}$. Parameters are taken the same as Fig. 1.}
\end{figure}

\begin{figure}[tbp]
\begin{center}
\includegraphics[scale=1.0]{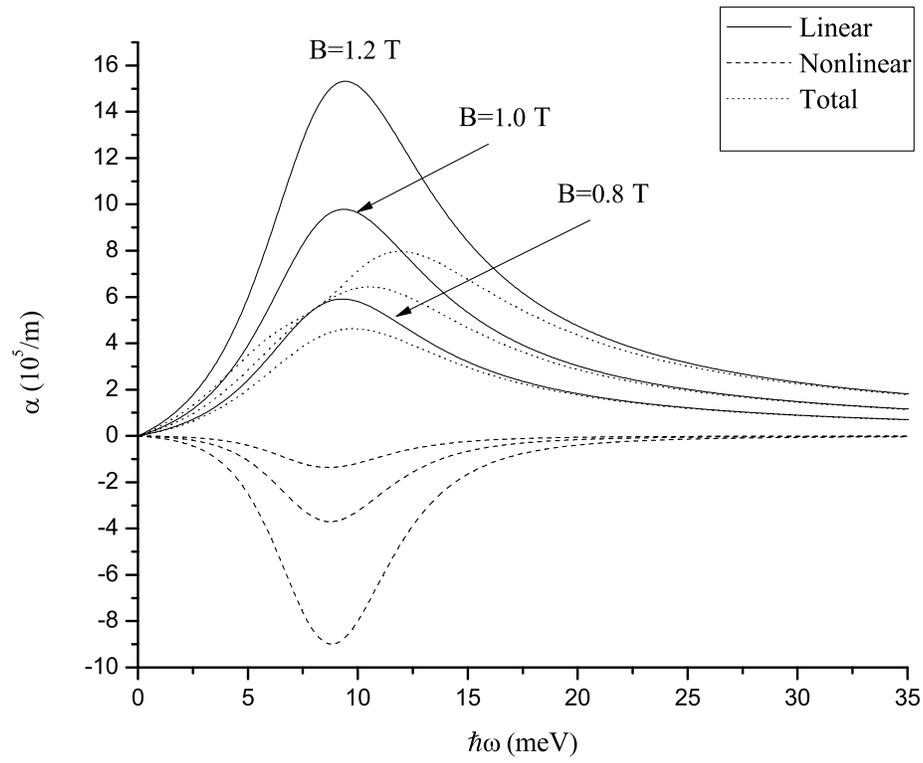}
\end{center}
\caption{The linear, the third-order nonlinear and the total ACs of a three-dimensional anisotropic QD as a function of the incident photon energy $\hbar\omega$ for three different values of the applied magnetic field $B$ with $\Omega_{x}=1.2\times10^{13} s^{-1}$, $I=1.0 Mw/cm^2$.}
\end{figure}

\begin{figure}[tbp]
\begin{center}
\includegraphics[scale=1.0]{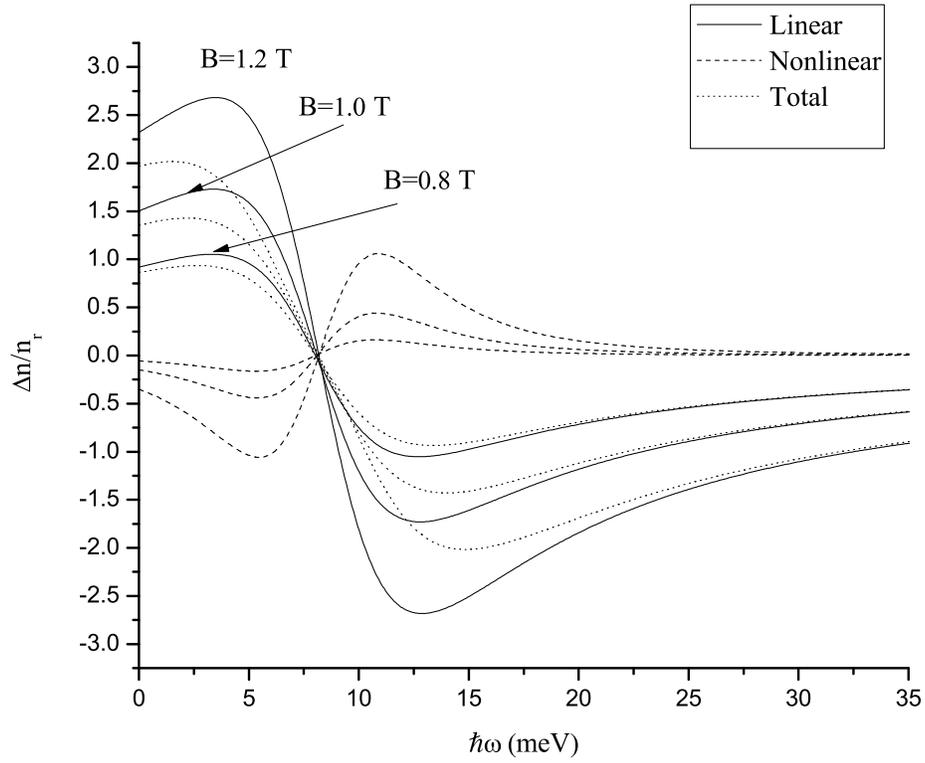}
\end{center}
\caption{The linear, the third-order nonlinear and the total RI changes of a three-dimensional anisotropic QD as a function of the incident photon energy $\hbar\omega$ for three different values of the applied magnetic field $B$. Parameters are taken the same as Fig. 3.}
\end{figure}

\begin{figure}[tbp]
\begin{center}
\includegraphics[scale=1.0]{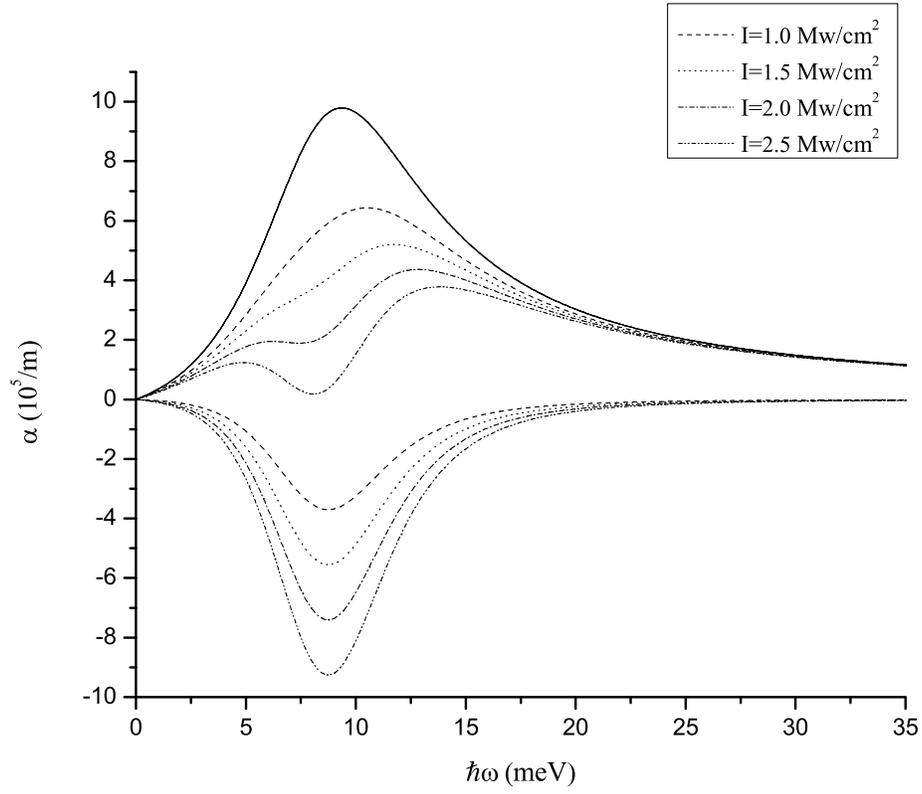}
\end{center}
\caption{The linear, the third-order nonlinear and the total ACs of a three-dimensional anisotropic QD as a function of the incident photon energy $\hbar\omega$ for three different values of the incident optical intensity $I$ with $\Omega_{x}=1.2\times10^{13} s^{-1}$, $B=1.0T$.}
\end{figure}

\begin{figure}[tbp]
\begin{center}
\includegraphics[scale=1.0]{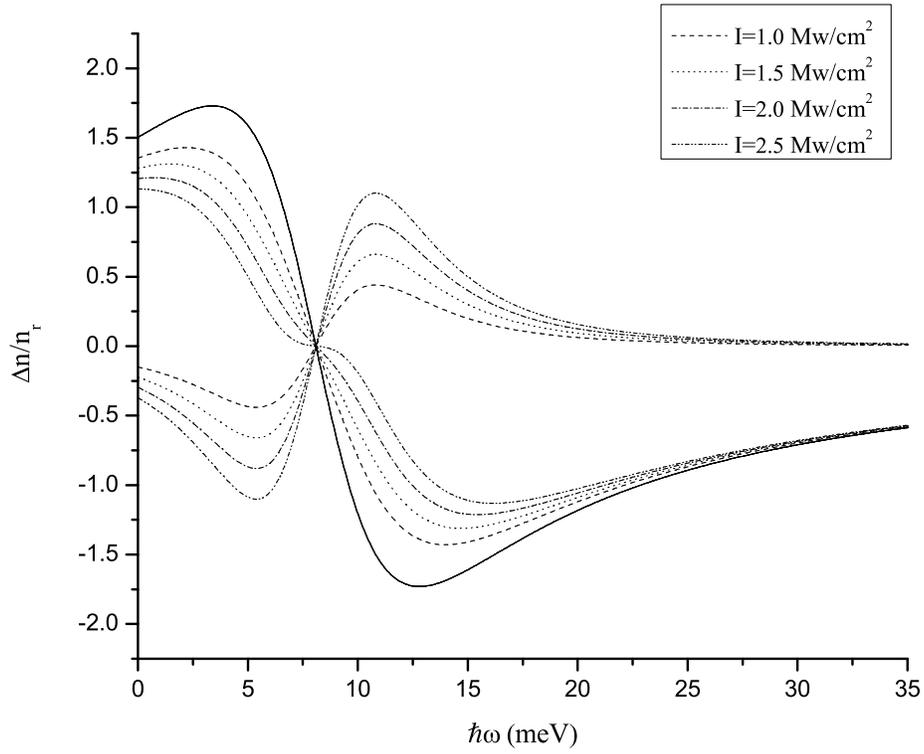}
\end{center}
\caption{The linear, the third-order nonlinear and the total RI changes of a three-dimensional anisotropic QD as a function of the incident photon energy $\hbar\omega$ for three different values of the incident optical intensity $I$. Parameters are taken the same as Fig. 5.}
\end{figure}

\begin{figure}[tbp]
\begin{center}
\includegraphics[scale=1.0]{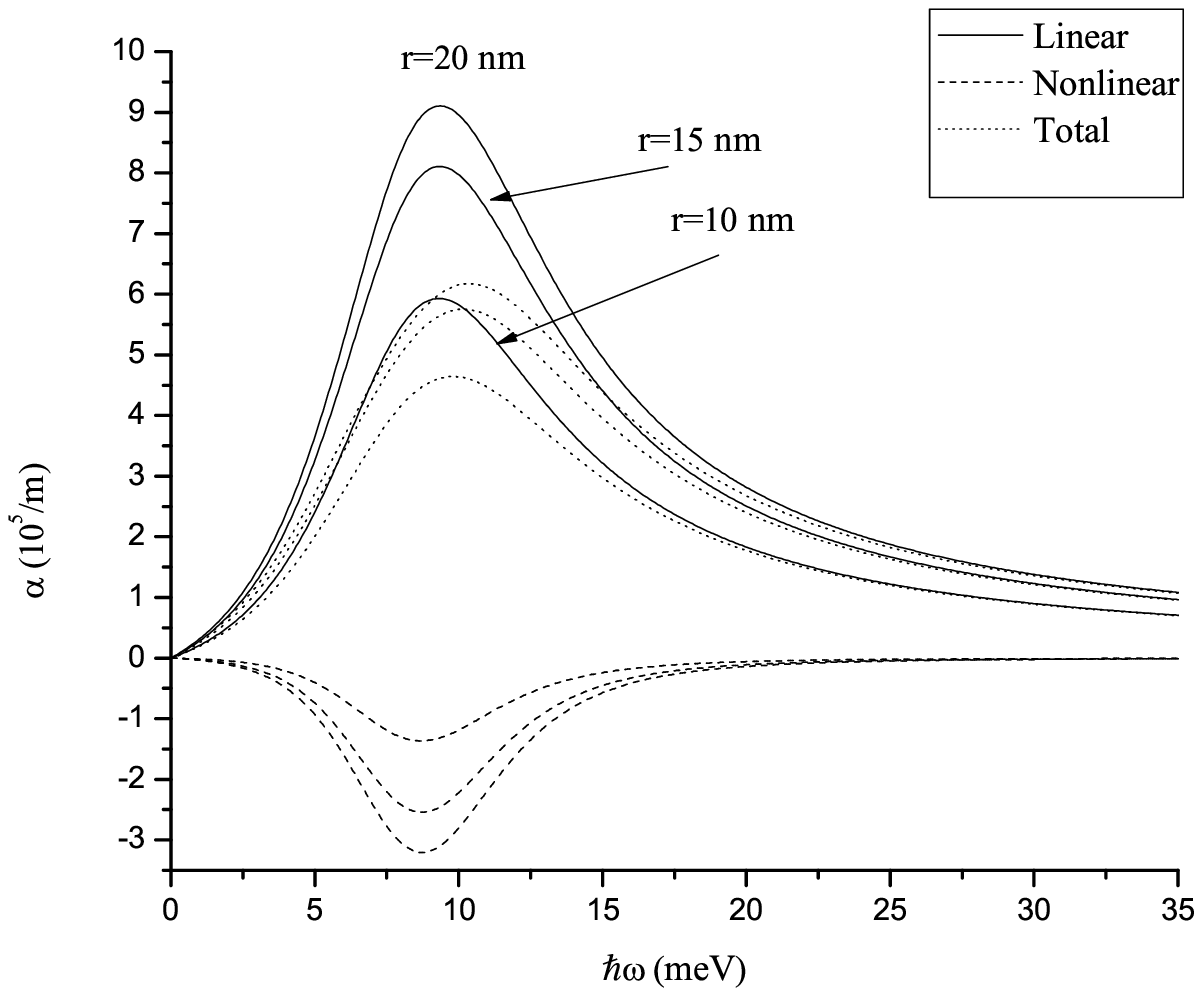}
\end{center}
\caption{The linear, the third-order nonlinear and the total ACs of a three-dimensional anisotropic QD as a function of the incident photon energy $\hbar\omega$ for three different values of the dot radius $r$ with $\Omega_{x}=1.2\times10^{13} s^{-1}$, $B=1.0 T$ and $I=1.0 Mw/cm^2$.}
\end{figure}

\begin{figure}[tbp]
\begin{center}
\includegraphics[scale=1.0]{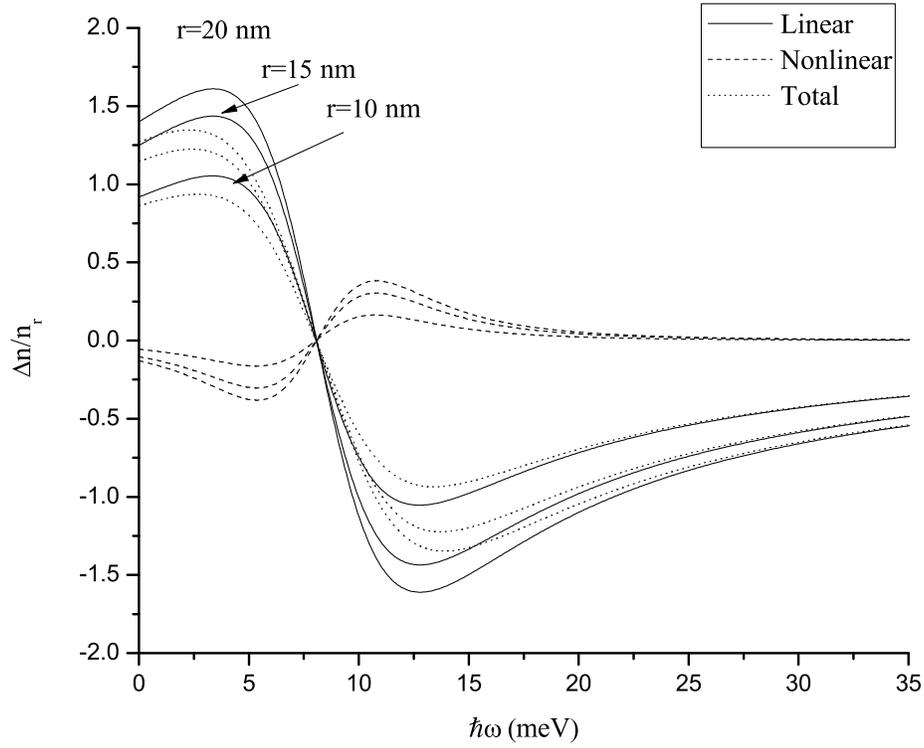}
\end{center}
\caption{The linear, the third-order nonlinear and the total RI changes of a three-dimensional anisotropic QD as a function of the incident photon energy $\hbar\omega$ for three different values of the dot radius $r$. Parameters are taken the same as Fig. 7.}
\end{figure}

\end{document}